\documentclass[useAMS,usenatbib,usegraphicx]{mn2e}

\newcommand{\etalb}{et al.}
\newcommand{\beq}{\begin{equation}}
\newcommand{\beqa}{\begin{eqnarray}}
\newcommand{\eeq}{\end{equation}}
\newcommand{\eeqa}{\end{eqnarray}}
\newcommand{\del}{\delta}

\title[On Correlated Random Walks and 21-cm Fluctuations During 
Cosmic Reionization] {On Correlated Random Walks and 21-cm
Fluctuations During Cosmic Reionization}

\author[R. Barkana]{R. Barkana
\thanks{E-mail: barkana@wise.tau.ac.il}\\
School of Physics and Astronomy, The Raymond and Beverly Sackler
Faculty of Exact Sciences, Tel Aviv University, Tel Aviv 69978,
ISRAEL}

\begin{document}

\pagerange{\pageref{firstpage}--\pageref{lastpage}} \pubyear{2006}

\maketitle

\label{firstpage}

\begin{abstract}
Analytical approaches to galaxy formation and reionization are based
on the mathematical problem of random walks with barriers. The
statistics of a single random walk can be used to calculate one-point
distributions ranging from the mass function of virialized halos to
the distribution of ionized bubble sizes during reionization. However,
an analytical calculation of two-point correlation functions or of
spatially-dependent feedback processes requires the joint statistics
of random walks at two different points. An accurate analytical
expression for the statistics of two correlated random walks has been
previously found only for the case of a constant barrier
height. However, calculating bubble sizes or accurate statistics for
halo formation involves more general barriers that can often be
approximated as linear barriers. We generalize the two-point solution
with constant barriers to linear barriers, and apply it as an
illustration to calculate the correlation function of cosmological
21-cm fluctuations during reionization.
\end{abstract}

\begin{keywords}
galaxies:high-redshift -- cosmology:theory -- galaxies:formation -- 
          large-scale structure of universe -- methods: analytical
\end{keywords}

\section{Introduction}

A critical prediction of any theory of structure formation is the mass
function of virialized dark-matter halos. While only numerical
simulations capture the full details of halo collapse, much of our
understanding of structure formation relies instead on analytical
techniques.  As such methods are based on simple assumptions and are
easily applied to a large range of models, they are indispensable both
for gaining physical understanding into the numerical results and
exploring the effects of model uncertainties. Analytical methods also
can be used to study and compensate for various limitations of
numerical simulations such as insufficient small-scale resolution,
missing large-scale fluctuations, or insufficiently early starting
redshifts.

The most widely applied method of this type was first developed by
\citet{ps74}. This simple model, later refined by \citet{bc91},
\citet{lc93}, and others, has had great success in describing the
formation of structure, reproducing rather accurately the numerical
results. Yet this model is intrinsically limited since it can only
predict the average number density of halos. Baryonic objects forming
within these halos are often subject to strong environmental effects
that are untreatable in this context. Many environmental effects such
as photoionization or metal enrichment are highly inhomogeneous in
nature, being caused by the nonlinear structures that form within the
intergalactic medium (IGM), and thus primarily impacting the areas
near these structures. Such interactions between the IGM and structure
formation are often better described as spatially-dependent feedback
loops rather than sudden changes in the overall average conditions.

The issue of spatial correlations also arises in another
context. Correlation functions are often an important statistic for
comparing theoretical predictions to the observed distribution of
objects. Some analytical models exist
\citep[e.g.,][]{kaiser84,ck89,mo96,st02} that supplement the
Press-Schechter number densities with additional approximate models,
but these do not arise naturally within the excursion set approach.

As we review in the following sections, a direct calculation of the
halo correlation function corresponds mathematically to solving for
the simultaneous evolution of two correlated random walks with a
constant barrier. This problem was first considered by \citet{p98},
who made some progress toward a satisfactory solution. We \citep{sb}
then found an approximate but quite accurate analytical solution and
used it to calculate the joint, bivariate mass function of halos
forming at two redshifts and separated by a fixed comoving
distance. We showed that our solution leads to a self-consistent
expression for the nonlinear biasing and correlation function of
halos, generalizing a number of previous results including those by
\citet{kaiser84} and \citet{mo96}. This solution has since been used 
to study, for example, the impact of clustered gas minihalos on cosmic
reionization \citep{bl02, iss05}, inhomogeneous metal enrichment at
high redshift \citep{ssf03}, and observations of metal lines around
Lyman break galaxies \citep{pm05}.

Recently, researchers have found useful applications for the more
general mathematical problem of random walks with a barrier that is
not constant. For instance, \citet{smt01} found that an ellipsoidal
collapse model suggests such a barrier for defining halos, yielding a
model that produces a halo mass function that better matches N-body
simulations. More recently, \citet{fzh04} used the statistics of a
random walk with a linear barrier to model the H II bubble size
distribution during the reionization epoch. While in principle this
distribution could be measured from maps of 21-cm emission by neutral
hydrogen, upcoming experiments such as the Mileura Widefield Array and
the Low Frequency Array are expected to be able to detect ionization
fluctuations only statistically, e.g., by measuring the correlation
function of the 21-cm brightness temperature \citep{miguel,CfA}. While
previously approximate expressions for spatial ionization correlations
have been developed \citep{fzh04, m05}, each of these was grafted
externally onto the underling formalism, requiring additional layers
of approximations.

In this paper we generalize the solution of \citet{sb} to linear
barriers and thus develop a self-consistent model for two-point
correlations in this case. The rest of this paper is organized as
follows. In \S~2 we establish our notation and review the simplest
case of the statistics of a single random walk with a constant
barrier. We then review in \S~3 the generalization to a single random
walk with a linear barrier. In \S~4 we follow the setup and solution of
\citet{sb} but generalize it to the case of two correlated random walks
with linear barriers. Since the barrier corresponding to the ionized
bubble size distribution during reionization is linear to a good
approximation, we use this distribution in \S~5 to illustrate how to
apply our results to explore various aspects of reionization and of
21-cm fluctuations that depend on two-point correlations among the
density and ionization fields. Our solution, however, is more general
and can be used in all problems where linear barriers are a good
approximation to the physical constraints. We briefly summarize our
results in \S~6.

\section{Single Random Walk With a Constant Barrier}

Before considering linear barriers, we first establish our notation
and review in this section the standard derivation of the one-point
expressions for a constant barrier within the context of the halo mass
function. The basic approach is that of \citet{bc91}, who rederived
and extended the halo formation model of \citet{ps74}.

We work with the linear overdensity field $\del({\bf x},z) \equiv
\rho({\bf x},z)/\bar\rho(z) - 1$, where ${\bf x}$ is a comoving
position in space, $z$ is the cosmological redshift and $\bar \rho$ is
the mean value of the mass density $\rho$. In the linear regime, the
overdensity grows in proportion to the linear growth factor $D(z)$
(defined relative to $z=0$). The barrier signifies the critical value
which this linearly-extrapolated $\delta$ must reach in order to
achieve some physical milestone; in the case of halo formation, an
estimate based on spherical top-hat collapse yields $\del_{c}=1.686$
\citep{p80} in the Einstein-de Sitter model.

A useful alternative parametrization is to consider the linear density
field extrapolated to the present time, i.e., the initial density
field at high redshift extrapolated to the present by multiplication
by the relative growth factor. In this case, the critical threshold
for collapse at redshift $z$ becomes redshift dependent even in the
Einstein-de Sitter case: \beq \del_c(z) = \del_{c} / D(z)\ .\eeq
However, it still represents a constant barrier at any given
redshift. We adopt this alternative view, and throughout this paper
the power spectrum $P(k)$ refers to the initial power spectrum,
linearly-extrapolated to the present (in particular, not including
non-linear evolution).

At a given $z$, we consider the smoothed density in a region around a
fixed point $A$ in space. We begin by averaging over a large mass
scale $M$, or, equivalently, by including only small comoving
wavenumbers $k$. We then lower $M$ until we find the highest value for
which the averaged overdensity is higher than $\del_c(z)$ and assume
that the point $A$ belongs to a halo with a mass $M$ corresponding to
this filter scale. In particular, if the initial density field is a
Gaussian random field and the smoothing is done using sharp $k$-space
filters, then the value of the smoothed $\del$ undergoes a random walk
as the cutoff value of $k$ is increased. If the random walk first hits
the collapse threshold $\del_c(z)$ at $k$, then at a redshift $z$ the
point $A$ is assumed to belong to a halo with a mass corresponding to
this value of $k$. Instead of using $k$, we adopt the variance as the
independent variable:
\beq
S_k \equiv \frac{1}{2 \pi^2} \int_0^k dk'\, k'^2\, P(k')\ .
\label{eq:Sk}
\eeq

In order to construct the number density of halos in this approach, we
need to find the probability distribution $Q(\del,S_k)$, where
$Q(\del,S_k)\, d\del$ is the probability for a given random walk to be
in the interval $\del$ to $\del+d\del$ at $S_k$. Alternatively,
$Q(\del,S_k)\, d\del$ can also be viewed as the trajectory density,
i.e., the fraction of the trajectories that are in the interval $\del$
to $\del+d\del$ at $S_k$, assuming that we consider a large ensemble
of random walks all of which begin with $\del=0$ at $S_k=0$.

The distribution $Q(\del,S_k)$ satisfies a diffusion equation
\beq
\frac{\partial Q}
{\partial S_k} = \frac{1} {2} \frac{\partial^2 Q} {\partial \del^2},
\label{eq:oneddiff}
\eeq
which is satisfied by the usual Gaussian solution:
\beq
G(\del,S_k) \equiv \frac{1}{\sqrt{2 \pi S_k}} \exp \left[
-\frac{\del^2}{2\, S_k} \right]\ .
\eeq

To determine the probability of halo collapse at a redshift $z$, we
consider random walks with an absorbing barrier at $\del=\nu$, where
for halo formation we set $\nu=\del_c(z)$. The solution with the
constant barrier in place is given by adding an extra image solution
\citep{ch43,bc91}:
\beq
Q_{\rm con}(\nu,\del,S_k)= G(\del,S_k)-G(2 \nu-\del,S_k)\ ,
\label{eq:const}
\eeq
where the subscript ``con'' refers to the constant barrier case. The
second (``image'') term is clearly (through a simple change of
variables) itself a solution to the diffusion equation, and the
combination $Q_{\rm con}$ is identically zero on the barrier
$\del=\nu$, hence it solves the diffusion equation and satisfies the
required boundary conditions.

The fraction of all trajectories that have hit the barrier $\nu$ by
$S_k$ includes all trajectories except those (represented by the
solution $Q_{\rm con}$) that still have not been absorbed:
\beq
F_{\rm >,con}(\nu,S_k) = 1-
\int_{-\infty}^{\nu}  d\del\, Q_{\rm con}(\nu,\del,S_k) =
2 \int_{\nu}^{\infty} d\del\, G(\del,S_k)\ .
\eeq
The differential of this is the first-crossing distribution:
\beqa
\lefteqn{f_{\rm con}(\nu,S_k) =
\frac{\partial }{\partial S_k} F_{\rm >,con}(\nu,S_k) = 
\left(\frac{\partial G(\del,S_k)}{\partial \del} \right)^
{\del=\infty}_{\del=\nu}} \nonumber \\ && \ \ \ \ \ \ \ \ =
\frac{\nu }{\sqrt{2 \pi} S_k^{3/2}} \exp\left[-\frac{\nu^2}
{2 S_k}\right]\ , \label{eq:f1pt} 
\eeqa 
where in the second equality we have used the fact that $G$ satisfies
eq.~(\ref{eq:oneddiff}). Note that $f(\nu,S_k)\, dS_k$ is the
probability that a random trajectory crosses the barrier in the
interval $S_k$ to $S_k + dS_k$.

In the halo interpretation, $f(\nu,S_k)\, dS_k$ is the probability
that a given point $A$ is in a halo with mass in the range
corresponding to $S_k$ to $S_k+d S_k$. The halo abundance is then
simply
\beq
\frac{dn}{dM} = \frac{\bar{\rho}_0}{M} \left|\frac{d S_k}{d M} \right|
f(\nu,S_k)\ ,
\label{eq:abundance}
\eeq
where $dn$ is the comoving number density of halos with masses in the
range $M$ to $M+dM$. The cumulative mass fraction in halos above mass
$M$ (thus denoted $F_>$) is similarly determined to be
\beq
F(>M | z) = F_{\rm >,con}(\nu,S_k) = {\rm erfc}
\left(\frac{\nu} {\sqrt{2 S_k}} \right)\ . \label{eq:ePS}
\eeq
Note that the complement of this is
\beq
F(<M | z) = F_{\rm <,con}(\nu,S_k) = {\rm erf}
\left(\frac{\nu} {\sqrt{2 S_k}} \right)\ .
\eeq

While these expressions were derived in reference to sharp $k$-space
smoothing (eq.~(\ref{eq:Sk})), $S_k$ is usually replaced in the final
results with the variance of the mass $M$ enclosed in a spatial sphere
of comoving radius $r$:
\beq
S_r(M) = S_r(r) = \frac{1}{2 \pi^2} \int_0^\infty k^2 dk\, P(k) W^2(k
r)\ ,
\label{eq:sig}
\eeq
where $W(x)$ is the spherical top-hat window function, defined in
Fourier space as
\beq
W(x) \equiv 3 \left[ \frac{\sin(x)}{x^3} - \frac{\cos(x)}{x^2} \right]\ .
\label{eq:reW}
\eeq 
The idea of this approach is that the real-space window function
corresponds more closely to spherical collapse (which yielded the
critical collapse threshold), while the mathematical problem is
simpler in $k$ space and leads to closed-form solutions.

\section{Single Random Walk With a Linear Barrier}

The problem of a random walk with a linear barrier has been previously
considered both in the context of improved halo mass functions and the
ionized bubble distribution. The problem is the same as that
considered in the previous section but with a barrier that is linear
in the variable $S_k$, i.e., has the form
\beq \delta = \nu + \mu S_k\ .
\eeq
The first-crossing distribution in this case was first derived by
\citet{s98}, while the full distribution function $Q(\del,S_k)$ in
this case was first worked out by \citet{m05}.

A key point that allows our further derivations below is that we find
a very simple way to express the solution of \citet{m05}:
\beq
Q_{\rm lin}(\nu,\mu,\del,S_k)= G(\del,S_k)-e^{-2 \nu \mu} G(2
\nu-\del,S_k)\ . \label{eq:lin}
\eeq
It is easy to check that this simple linear modification of the usual
image solution of eq.~(\ref{eq:const}) is identically zero on the
linear barrier, as required. 

Integrating, we find the fraction of all trajectories that have
reached the barrier by $S_k$:
\beqa
\lefteqn{ F_{\rm >,lin}(\nu,\mu,S_k) = 1-
\int_{-\infty}^{\nu+\mu S_k}  d\del\, Q_{\rm lin}(\nu,\mu,\del,S_k)} 
\label{eq:Flin} \\ & & \ \ \ \ \ = \frac{1}{2} \left[ {\rm erfc}
\left(\frac{\nu+\mu S_k} {\sqrt{2 S_k}} \right) + e^{-2 \nu \mu}\, 
{\rm erfc}\left(\frac{\nu-\mu S_k} {\sqrt{2 S_k}} \right) \right]\
. \nonumber
\eeqa
This expression agrees with that in \citet{m05}. The complement is
\beqa
\lefteqn{ F_{\rm <,lin}(\nu,\mu,S_k) = 1-
F_{\rm >,lin}(\nu,\mu,S_k)} \\ & & 
\ \ \ \ \ = \frac{1}{2} \left[ {\rm erfc}\left(\frac{-\nu-\mu S_k}
{\sqrt{2 S_k}} \right) - e^{-2 \nu \mu}\, {\rm
erfc}\left(\frac{\nu-\mu S_k} {\sqrt{2 S_k}} \right) \right]\
. \nonumber
\eeqa
Also needed for later is the first moment of the density among
trajectories that do not hit the barrier:
\beqa
\lefteqn{\bar{\delta}_{\rm lin}(\nu,\mu,S_k)= \int_{-\infty}
^{\nu+\mu S_k} d\del\, \del\, Q_{\rm lin} (\nu,\mu,\del,S_k)}
\nonumber \\ & & \ \ \ \ \ \ \ \ \ = -\nu e^{-2 \nu
\mu}\, {\rm erfc}\left(\frac{\nu-\mu S_k} {\sqrt{2 S_k}} \right)\ .
\label{eq:dbar} \eeqa
Finally, we differentiate to obtain the first-crossing distribution in
agreement with \citet{s98}:
\beqa
\lefteqn{
f_{\rm lin}(\nu,\mu,S_k) = \frac{\partial }{\partial S_k} F_{\rm
>,lin}(\nu,\mu,S_k)} \nonumber \\ && \ \ \ \ \ \ \ \ \ = \frac{\nu
}{\sqrt{2 \pi} S_k^{3/2}} \exp\left[-\frac{(\nu+\mu S_k)^2} {2
S_k}\right]\ .
\eeqa

\section{Two Correlated Random Walks With Linear Barriers}

\subsection{Analytic Preliminaries}

We follow \citet{sb} in setting up the problem of the statistics of
two correlated random walks. We consider points $A$ and $B$ separated
by a fixed comoving distance $d$. Note that this definition of
distance is in Lagrangian space, which is intrinsic to any
Press-Schechter type approach. Thus, it is the comoving distance
between points $A$ and $B$ at early times, and does not take into
account subsequent peculiar motions of these points. If we consider
smoothed densities identified by sharp $k$-space filters $k_1$ at
point $A$ and $k_2$ at point $B$, then the cross-correlation of the
densities involves only those $k$ values common to both filters, and
its value is
\beq \xi_k(d,S_k) \equiv \frac{1}{2 \pi^2} \int_0^{k} k'^2\, dk'\,
\frac{\sin(k'd)} {k'd}\, P(k')\ ,
\label{eq:xik}
\eeq
where the upper integration limit is $k={\rm min}[k_1,k_2]$, and when
we write $\xi_k$ as a function of $S_k$ it is related to $k$ by
eq.~(\ref{eq:Sk}). It is also convenient to define
\beq
\eta(d,S_k)
\equiv \frac{\sin\left[k(S_k)\, d\right]}{k(S_k)\, d}\ ,
\label{eq:etak}
\eeq
so that \beq
\xi_k (d,S_k)= \int_{S'=0}^{S_k} \eta(d,S_k')\, dS_k'\ .  \label{eq:xi}
\eeq

Just as the real-space variance is often used in the one-point case,
the two-point quantities we discuss below will use the correlation
between two spatial filters centered about two points at a separation
$d$. In this case, the standard expression is
\beqa
\lefteqn{\xi_r(d,r_1,r_2) \equiv \frac{1}{2 \pi^2} \int_0^\infty k^2 
dk\, \frac{\sin(k d)}{k d} P(k) W(k r_1) W(k r_2)\ ,} \nonumber \\
\label{eq:xi2}
\eeqa
where $r_1$ and $r_2$ are the radii of the two filters, and $W(x)$ is
again the top-hat window function given by eq.~(\ref{eq:reW}).
However, \citet{sb} showed that in order to ensure the most
physically-reasonable behavior of the solution in various regions of
the parameter space (including in limits that reduce to the one-point
case), it is better to substitute for $\xi_k$ the quantity
\beq
\xi_{r_{max}}(d,r_1,r_2)
\equiv \xi_r[d,{\rm max}(r_1,r_2), {\rm max}(r_1,r_2)]\ ,
\label{eq:xirmax}
\eeq
which is equal to $\xi_r$ when the two filters have equal radii. When
the variances $S_1$ and $S_2$ are used as the fundamental variables,
we find the corresponding $r_1$ and $r_2$ by inverting the relation
$S_r(r)$ given by eq.~(\ref{eq:sig}).

\subsection{Basic Setup}

We continue to follow \citet{sb} as we consider simultaneous
correlated random walks of two overdensities $\del_1(S_{k,1})$ and
$\del_2(S_{k,2})$ separated by a fixed Lagrangian distance $d$. As in
the one-point case, for the derivation we adopt sharp $k$-space
filters. We want to determine the joint probability distribution of
these two densities, $Q(\del_1,\del_2,S_{k,1},S_{k,2},d).$ In terms of
a trajectory density in the $(\del_1,\del_2)$ plane,
$Q(\del_1,\del_2,S_{k,1},S_{k,2},d)\, d\del_1\, d\del_2$ is the
fraction of trajectories that are in the interval $\del_1$ to
$\del_1+d\del_1$ and $\del_2$ to $\del_2+d\del_2$ at $(S_{k,1},
S_{k,2})$.  Below we will take $S_{k,1}$ and $S_{k,2}$ to be the {\em
final}\, variances of these trajectories, denoting intermediate
variances with the primed notation $S'_{k,1}$ and $S'_{k,2}.$ We then
consider a large number of random walks all of which begin with
$\del_1=0$ and $\del_2=0$ at $S'_{k,1}=0$ and $S'_{k,2}=0$.

With sharp $k$-space filters, the problem simplifies due to the fact
that we are working with a Gaussian random field. \citet{sb} showed
that we can consider $Q$ to be a function of a single variable $S'_k$,
with a diffusion equation
\beqa
\lefteqn{
\frac{\partial Q} {\partial S'_k} =
\cases {
\frac{1} {2}  \frac{\partial^2 Q} {\partial \del_1^2} +
\eta(d,S'_k)\,
\frac{\partial^2 Q} {\partial \del_1 \del_2}+
\frac{1}{2} \frac{\partial^2 Q} {\partial \del_2^2}
	& $\ S'_k < S_{k,{\rm min}}$ \cr
\frac{1}{2}  \frac{\partial^2 Q} {\partial \del_1^2}
	& $\ S_{k,2} < S'_k < S_{k,1}$ \cr
\frac{1}{2}  \frac{\partial^2 Q} {\partial \del_2^2}
	& $\ S_{k,1} < S'_k < S_{k,2}$\ ,
\cr} } \nonumber \\
\eeqa
where $S_{k,{\rm min}}$ is the smaller of $S_{k,1}$ and $S_{k,2}$.

\subsection{Two-Step Approximation} 

While the full solution of the double barrier problem requires a
numerical approach, \citet{sb} found a simple approximate analytic
solution that captures the underlying physics of two-point collapse.

Consider the expression for the differential correlation coefficient
$\eta(d,S'_k)$, eq.~(\ref{eq:etak}). While this is an oscillating
function, it equals unity at small values of $S'_k$ and its amplitude
declines towards zero once $k d \gg 1$. Thus, for small $S'_k$ values,
the two random walks are essentially identical, while at large $S'_k$,
the two random walks become independent.

These observations led \citet{sb} to propose a ``two-step''
approximation in which $\eta(d,S'_k)$ is replaced with a simple step
function. In order to preserve the exact solution for $Q$ at
$S'_k=S_{k,{\rm min}}$ in the absence of the barriers, we specifically
took
\beq
\eta(d,S'_k) \simeq \cases { 1 & $0 \leq
S'_k \leq \xi_k(d,S_{k,{\rm min}})$ \cr 
0 & $\xi_k(d,S_{k,{\rm min}}) < S'_k
\leq S_{k,\rm min} $\ . \cr} 
\eeq

Hereafter we adopt a general notation for the variances and
correlation functions, using $S$ to represent either the $k$-space
filtered quantity, $S_k$, its real space equivalent $S_r$, or any
alternative definition. Similarly, $\xi$ denotes $\xi_{r_{max}}$ or
$\xi_{k(r)}$. Following the common approximation taken in the
single-particle case, in all applications we use the real-space
quantities, i.e., $S_1$ and $S_2$ denote $S_r(M_1)$ and $S_r(M_2)$,
respectively. Also, we use for $\xi$ the correlation function
$\xi_{r_{max}}$ as given by eq.~(\ref{eq:xirmax}). Note that although
we write the dependence of various functions on $\xi$ explicitly,
$\xi$ is not an independent variable but instead is a function of
$S_1$ and $S_2$ (as well as the separation $d$).

\subsection{Analytic Solution With Linear Barriers}

Using the two-step approximation, \citet{sb} found the analytic
solution for $Q$ in the case of constant absorbing barriers at
$\del_1=\nu_1$ and $\del_2=\nu_2$. We follow their derivation, but
generalize it to the case of two (possibly different) linear barriers
at $\del_1=\nu_1+\mu_1 S'$ and $\del_2=\nu_2+\mu_2 S'$. Under the
two-step approximation, we must first evolve $\del_1$ for $0 \leq S'
\leq \xi$. Since we are assuming that the two random walks are
identical in this regime, we must place the barrier on $\del_1$ at
$\del_1=\nu_{\rm m}+\mu_{\rm m} S'$ where we choose $\nu_{\rm m}$ and
$\mu_{\rm m}$ to be as large as possible such that the resulting
barrier still lies below both of the original linear barriers,
throughout the relevant range of $S'$. In principle, the best
approximation would be to adopt at each $S'$ the lower of the two
barriers. However, if the barriers were to cross within the range $0
\leq S' \leq \xi$, this would require an extra convolution compared to
our solution below and would thus complicate it
substantially. Fortunately, this appears not to be needed in practice,
at least in our main application which is reionization. In the
examples given in \S~5, we find that while $\nu$ changes rapidly with
redshift, $\mu$ (which does change in the opposite direction) varies
extremely slowly, so that if barriers are considered at two different
redshifts, the lower-redshift one is the lower barrier at all relevant
values of $S'$.

Quantitatively, the solution for a single linear absorbing barrier,
eq.~(\ref{eq:lin}), gives $Q$ at $S'=\xi$:
\beqa
\lefteqn{Q_a(\nu_{\rm m},\mu_{\rm m},\del_1,\del_2, \xi) =} \nonumber 
\\ && \left[ G(\del_1, \xi)- e^{-2 \nu_{\rm m} \mu_{\rm m}} G(2
\nu_{\rm m} - \del_1, \xi)\right] \nonumber \\ && \mbox{} \times
\del_D(\del_1 - \del_2) \theta(\nu_{\rm m} +
\mu_{\rm m} \xi - \delta_1)\ , \label{eq:Qa}
\eeqa
where $\del_D$ is a one-dimensional Dirac delta function and $\theta$
is the Heaviside step function. We then evolve the random walks in
$\del_1$ and $\del_2$ independently from their common starting point
at $\xi$ up to $S_1$ and $S_2$, with the barriers at $\nu_1+\mu_1 S'$
and $\nu_2+\mu_2 S'$, respectively.  Thus, we first convolve
eq.~(\ref{eq:Qa}) with the no-barrier solutions for the two
independent random walks,
\beqa
\lefteqn{Q_{b}(\del_1,\del_2,S_1,S_2, \xi) = 
G(\del_1,S_1- \xi)\, G(\del_2,S_2 - \xi)\ .} \nonumber \\
\eeqa
Letting $\delta$ be the value of $\del_1$ at $S'=\xi$, we can
write this convolution explicitly as
\beqa
\lefteqn{
Q_0(\nu_{\rm m},\mu_{\rm m},\del_1,\del_2,S_1,S_2,\xi)=}
\label{eq:conv}\\ \lefteqn{ \int_{\delta=-\infty}^{\nu_{\rm m}+
\mu_{\rm m} \xi} d \del\  
Q_{\rm lin}(\nu_{\rm m},\mu_{\rm m},\del,\xi) } \nonumber 
\\ && \ \ \ \ \ \ \ \mbox{} \times G(\del_1-\del,S_1- \xi)\, 
G(\del_2-\del, S_2 - \xi)\ . \nonumber
\eeqa
Evaluating this yields
\beqa
\lefteqn{
Q_0(\nu_{\rm m},\mu_{\rm m},\del_1,\del_2,S_1,S_2,\xi)=}
\label{eq:q0two} \\ && Q_+(\nu_{\rm m}+\mu_{\rm m}
\xi,\del_1,\del_2,S_1,S_2,\xi) \nonumber \\ && 
\mbox{} + e^{-2 \nu_{\rm m} \mu_{\rm m}} \nonumber \\ && \ \ 
\mbox{} \times Q_-(\nu_{\rm m}-\mu_{\rm m} \xi,2
\nu_{\rm m} - \del_1, 2 \nu_{\rm m} - \del_2,S_1,S_2,
\xi)\ , \nonumber 
\eeqa
where
\beqa
\lefteqn{Q_{\pm} (\nu,\del_1,\del_2,S_1,S_2,\xi)} 
\nonumber \\  & & \equiv \frac{1}{4 \pi \sqrt{S_1 S_2 -
\xi^2}} \nonumber \\ & & \mbox{} \times \exp \left[ -\, 
\frac{\del_1^2 S_2 + \del_2^2 S_1 - 2 \del_1 \del_2 \xi} 
{2 (S_1 S_2 - \xi^2)} \right]  \nonumber \\
& & \mbox{} \times \left[{\rm erf} \left( \tilde \nu
\sqrt{ \frac{\tilde S}{2}}\right) \pm 1 \right]\ ,
\label{eq:Qpm}
\eeqa
and we have defined
\beqa
\tilde S & \equiv & \frac{ \xi (S_1 - \xi) (S_2 - \xi)}
{S_1 S_2 - \xi^2}\ , \nonumber \\ \tilde \nu & \equiv & \
\frac{\nu}{\tilde S} - \frac{\del_1}{S_1 - \xi} -
\frac{\del_2}{S_2 - \xi}\ .
\eeqa
Note that the quantities $Q_{\pm}$ are unchanged from \cite{sb}, but
the solution for $Q_0$ is now more general.

Finally, we must account for the additional barriers on $\del_1$ and
$\del_2$ in the regime where their random walks are independent.  To
do this, we first note that the barrier $\del_1 = \nu_1+\mu_1 S'$ can
be written also as a linear barrier in terms of the relative variables
$\del_1-\del$ and $S' - \xi$: the barrier is at $\del_1 - \del = [
\nu_1+\mu_1 \xi - \del] + \mu_1 (S' - \xi)$. Thus, the linear-barrier
solution of eq.~(\ref{eq:lin}) shows that we must subtract from the
no-barrier term $G(\del_1-\del,S_1-
\xi)$ in eq.~(\ref{eq:conv}) an image-like second term: 
\beq
e^{-2 (\nu_1 + \mu_1 \xi - \del) \mu_1}\ G(2 (\nu_1 + \mu_1 \xi -
\del) -(\del_1-\del),S_1- \xi)\ . \label{eq:diff}
\eeq
Thus, the solution $Q$ can be written as
\beqa
\lefteqn{
Q(\nu_1,\nu_2,\mu_1,\mu_2,\del_1,\del_2,S_1,S_2,\xi)=}
\nonumber\\ \lefteqn{ \int_{\delta=-\infty}^{\nu_{\rm m}+
\mu_{\rm m} \xi} d \del\ 
Q_{\rm lin}(\nu_{\rm m},\mu_{\rm m},\del,\xi) } \nonumber 
\\ && \ \ \ \ \ \ \ \mbox{} \times Q_{\rm lin}(\nu_1+\mu_1 \xi - \del,
\mu_1,\del_1 - \del,S_1 - \xi)\, \nonumber \\ 
&& \ \ \ \ \ \ \ \mbox{} \times Q_{\rm lin}(\nu_2+\mu_2 \xi - \del,
\mu_2,\del_2 - \del,S_2 - \xi)\ . \label{eq:conv1}
\eeqa

This integral is difficult since the exponential factor that
multiplies $G$ in eq.~(\ref{eq:diff}) itself contains the integration
variable $\del$, and the result therefore cannot immediately be
written in terms of $Q_0$. However, we solve this difficulty by noting
that the expression in eq.~(\ref{eq:diff}) can be written in an
equivalent, alternate form:
\beq
e^{-2 (\del_1 - (\nu_1 + \mu_1 S_1)) \mu_1}\ G(\del_1 + \del - 2
(\nu_1 + \mu_1 S_1) ,S_1- \xi)\ .\label{eq:diff2}
\eeq
Thus, the solution $Q$ can also be written as
\beqa
\lefteqn{
Q(\nu_1,\nu_2,\mu_1,\mu_2,\del_1,\del_2,S_1,S_2,\xi)=}
\nonumber\\ \lefteqn{ \int_{\delta=-\infty}^{\nu_{\rm m}+
\mu_{\rm m} \xi} d \del\ 
Q_{\rm lin}(\nu_{\rm m},\mu_{\rm m},\del,\xi) } \nonumber 
\\ && \ \ \ \ \ \ \ \mbox{} \times Q_{\rm lin}(\del_1 - (\nu_1 + \mu_1 S_1),
\mu_1,\del_1 - \del,S_1 - \xi)\, \nonumber \\ 
&& \ \ \ \ \ \ \ \mbox{} \times Q_{\rm lin}(\del_2 - (\nu_2 + \mu_2 S_2),
\mu_2,\del_2 - \del,S_2 - \xi)\ . \label{eq:conv2}
\eeqa
This integration yields the complete solution:
\beqa
\lefteqn{
Q(\nu_1,\nu_2,\mu_1,\mu_2,\del_1,\del_2,S_1,S_2,\xi)=}
\nonumber \\ && Q_0(\nu_{\rm m},\mu_{\rm m},\del_1,\del_2,S_1,
S_2,\xi) \nonumber \\ && \mbox{} + \exp \left[2 (\del_1^{\rm br} 
- \del_1) \mu_1+ 2 (\del_2^{\rm br} - \del_2) \mu_2 \right] 
\nonumber \\ && \ \ \mbox{} \times Q_0(\nu_{\rm m},\mu_{\rm m},2 
\del_1^{\rm br}-\del_1,2 \del_2^{\rm br}-\del_2,S_1,S_2,
\xi) \nonumber \\ && \mbox{} - \exp \left[2 (\del_2^{\rm br} - 
\del_2)\mu_2\right] \nonumber \\ && \ \ \mbox{} \times Q_0(\nu_{\rm m},
\mu_{\rm m}, \del_1,2 \del_2^{\rm br} -\del_2,S_1,S_2,\xi)
\nonumber \\ && \mbox{} - \exp \left[2 (\del_1^{\rm br} - \del_1) 
\mu_1 \right] \nonumber \\ && \ \ \mbox{} \times Q_0(\nu_{\rm m},
\mu_{\rm m},2 \del_1^{\rm br}-\del_1, \del_2,S_1,S_2,\xi)\ ,
\label{eq:soln}
\eeqa
where we have defined the $\del$ values on the barriers:
\beq \del_1^{\rm br} \equiv \nu_1 + \mu_1 S_1;\ \ \ \ \ \ 
\del_2^{\rm br} \equiv \nu_2 + \mu_2 S_2\ .
\eeq

\cite{sb} showed that the solution with the two-step approximation 
is very accurate in the case of constant barriers, giving the same
results as a full numerical solution to within at most $2\%$, with the
difference typically much smaller than this value. Since the idea of
the approximation (as presented in the previous subsection) is based
on the properties of two correlated walks and not on any particular
property of the barriers, we expect this approximation to be accurate
in the case of linear barriers as well.

\subsection{Bivariate Cumulative Distribution}

Having developed in the previous subsection an accurate approximation
to the joint statistics of two correlated random walks, we now apply
this distribution to find the joint probability of having the two
random walks cross their respective barriers before reaching two given
values $S_1$ and $S_2$. In particular applications, this quantity can
be interpreted as the joint probability that point $A$ is in a halo
above a mass $M_1(S_1)$ and point $B$ is in a halo above a mass
$M_2(S_2)$, or as the joint probability that point $A$ is in an
ionized bubble above some size (see \S~5.1) and point $B$ is in an
ionized bubble above some other given size.

Consider first the following quantity:
\beqa
\lefteqn{
 F_<(\nu_1,\nu_2,\mu_1,\mu_2,S_1,S_2,\xi) =} \nonumber \\ 
\lefteqn{\int_{-\infty}^{\nu_1+\mu_1 S_1} d\del_1 \int_{-\infty}^{\nu_2 
+ \mu_2 S_2} d\del_2\ Q(\nu_1,\nu_2,\mu_1,\mu_2,\del_1,\del_2,S_1,
S_2,\xi)\ .} \nonumber \\ \eeqa This is the probability that both
random walks are not absorbed before reaching the point $(S_1,
S_2)$. We denote it $F_<$ since, e.g., in the halo-formation case it
is the chance that point $A$ is in a halo below a mass $M_1(S_1)$ and
point $B$ is in a halo below a mass $M_2(S_2)$,

We can find an expression for $F_<$ in terms of a single integral (as
did \cite{sb} in the constant-barrier case), by writing $Q$ in the
form of eq.~(\ref{eq:conv1}) and performing the $\del_1$ and $\del_2$
integrals. The result is
\beqa
\lefteqn{
 F_<(\nu_1,\nu_2,\mu_1,\mu_2,S_1,S_2,\xi) =} \nonumber \\ 
\lefteqn{ \int_{\delta=-\infty}^{\nu_{\rm m}+ \mu_{\rm m} \xi} 
d \del\ Q_{\rm lin}(\nu_{\rm m},\mu_{\rm m},\del,\xi) } 
\nonumber \\ && \ \ \ \ \ \ \ \mbox{} \times F_{\rm <,lin}(\nu_1+\mu_1 
\xi - \del,\mu_1,S_1 - \xi) \nonumber \\ && \ \ \ \ \ \ \ \mbox{} 
\times F_{\rm <,lin}(\nu_2+\mu_2 \xi - \del,\mu_2,S_2 - \xi)\ .
\label{eq:F<}
\eeqa
This expression is easy to understand: After the correlated random
walk reaches $\del$ at $S' = \xi$, without hitting the joint barrier
$\nu_{\rm m}+ \mu_{\rm m} S'$, the subsequent random walks are
independent. We therefore multiply the probability that random walk
\#1 does not hit its barrier between $S' = \xi$ and $S_1$, with 
the probability that random walk \#2 does not hit its barrier between
$S' = \xi$ and $S_2$. This is then integrated over the probability
distribution of reaching various values of $\del$ at $S' = \xi$.

The complementary quantity $F_>$ is the probability that both random
walks are absorbed before reaching the point $(S_1, S_2)$. This cannot
be calculated with a similar expression as in eq.~(\ref{eq:F<}), just
replacing $F_{\rm <,lin}$ with $F_{\rm >,lin}$ in the integrand, since
the barrier can also be crossed before the correlated walk reaches $S'
= \xi$. Instead, we find $F_>$ as the complement of the chance that at
least one of the random walks is not absorbed. The latter chance
equals the chance that \#1 is not absorbed, plus the chance that \#2
is not absorbed, minus (to eliminate double counting) the chance that
both are not absorbed. We obtain from this:
\beqa
\lefteqn{
 F_>(\nu_1,\nu_2,\mu_1,\mu_2,S_1,S_2,\xi) =} \nonumber \\ 
&& 1 + F_<(\nu_1,\nu_2,\mu_1,\mu_2,S_1,S_2,\xi)
\nonumber \\ && \mbox{} - F_{\rm <,lin}(\nu_1,\mu_1,S_1)
- F_{\rm <,lin}(\nu_2,\mu_2,S_2) \ .
\label{eq:F>}
\eeqa
We can similarly calculate the mixed quantities; e.g., the chance that
walk \#1 is not absorbed but \#2 is absorbed is
\beqa
\lefteqn{
 F_{<>}(\nu_1,\nu_2,\mu_1,\mu_2,S_1,S_2,\xi) =} \nonumber \\ 
&& F_{\rm >,lin}(\nu_2,\mu_2,S_2) - 
F_>(\nu_1,\nu_2,\mu_1,\mu_2,S_1,S_2,\xi)\ .
\label{eq:F<>}
\eeqa
Finally, the chance that walk \#1 is absorbed but \#2 is not absorbed
is obtained by switching the indices 1 and 2 in eq.~(\ref{eq:F<>}).

\subsection{Other Distributions}

The bivariate first-crossing distribution $f\, dS_1\, dS_2$ is the
probability of having random walk \#1 cross the barrier in the range
$S_1$ to $S_1+d S_1$ and point \#2 cross in the range $S_2$ to $S_2+d
S_2$. This is simply related to the bivariate cumulative distribution
as
\beqa
\lefteqn{ f(\nu_1,\nu_2,\mu_1,\mu_2,S_1,S_2,\xi) =} \nonumber \\
&& \frac{\partial} {\partial S_1} \frac{\partial}{\partial S_2}  
F_<(\nu_1,\nu_2,\mu_1,\mu_2,S_1,S_2,\xi)\ , \label{eq:f}
\eeqa
where $\xi$ is not considered an independent variable (and so the
partial derivatives involve variations of $\xi$). The expression for
$f$ can in principle be simplified by bringing the derivatives inside
the integral in eq.~(\ref{eq:F<}) and using the properties of the
integrand as in the analogous case in \citet{sb}, although here the
expressions are more complicated (and there is also a contribution
from the $\xi$ that appears in the integration limit). Since we are
not directly interested in $f$ in the context of upcoming probes of
reionization, we do not develop this further here. The derivatives in
eq.~(\ref{eq:f}) can also be evaluated numerically.

Various correlated distributions of density and of ionization (i.e.,
hitting the barrier) can also be calculated with our solution. For
example, consider the probability distribution of $\delta_1$ at $S_1$
given that random walk \#1 has not been absorbed by its barrier while
\#2 has been absorbed by its barrier before $S_2$. We calculate this
as follows: After the correlated random walk reaches $\del$ at $S' =
\xi$, without hitting the joint barrier $\nu_{\rm m}+ \mu_{\rm m} S'$
(so that \#1 will be unabsorbed), the subsequent random walks are
independent. We therefore multiply the probability that random walk
\#1 reaches $\del_1$ at $S_1$ without hitting its barrier on the way,
with the probability that random walk
\#2 does hit its barrier between $S' = \xi$ and $S_2$. This is then
integrated over the probability distribution of reaching various
values of $\del$ at $S' = \xi$. The probability is proportional to the
following quantity:
\beqa
\lefteqn{
f_{[\del | <>]}(\nu_1,\nu_2,\mu_1,\mu_2,\del_1,S_1,S_2,\xi) =}
\nonumber \\ \lefteqn{ \int_{\delta=-\infty}^{\nu_{\rm m}+ \mu_{\rm m}
\xi} d \del\ Q_{\rm lin}(\nu_{\rm m},\mu_{\rm m},\del,\xi) } \nonumber
\\ && \ \ \ \ \ \ \ \mbox{} \times Q_{\rm lin}(\del_1 - (\nu_1 + \mu_1
S_1), \mu_1,\del_1 - \del,S_1 - \xi) \nonumber \\ && \ \ \ \ \ \ \
\mbox{} \times F_{\rm >,lin}(\nu_2+\mu_2 \xi - \del,\mu_2,S_2 - \xi) \
. \label{eq:pdel1}
\eeqa
As written, this distribution for $\del_1$ is not normalized; a
normalized probability distribution can be obtained by dividing by the
probability in eq.~(\ref{eq:F<>}) (with indices switched) that walk
\#1 is not absorbed by its barrier while \#2 is absorbed.

Finally, consider the probability distribution of $\del_1$ at $S_1$
given that both walks have not been absorbed. This is similar to
eq.~(\ref{eq:F<}) except that we integrate only over the values of
$\del_2$. Thus, we first calculate the quantity:
\beqa
\lefteqn{
f_{[\del | <]}(\nu_1,\nu_2,\mu_1,\mu_2,\del_1,S_1,S_2,\xi) =}
\nonumber \\ \lefteqn{ \int_{\delta=-\infty}^{\nu_{\rm m}+ \mu_{\rm m}
\xi} d \del\ Q_{\rm lin}(\nu_{\rm m},\mu_{\rm m},\del,\xi) } \nonumber
\\ && \ \ \ \ \ \ \ \mbox{} \times Q_{\rm lin}(\del_1 - (\nu_1 + \mu_1
S_1), \mu_1,\del_1 - \del,S_1 - \xi) \nonumber \\ && \ \ \ \ \ \ \
\mbox{} \times F_{\rm <,lin}(\nu_2+\mu_2 \xi - \del,\mu_2,S_2 - \xi) \
. \label{eq:pdel2}
\eeqa
A normalized probability distribution can be obtained from this by
dividing by the probability in eq.~(\ref{eq:F<}) that both walks are
not absorbed.

\section{Illustration: Cosmic Reionization}

\subsection{The Density and Ionization Fields}

We illustrate the application of our solution to reionization using
the model of \citet{fzh04} for the ionized bubble
distribution. According to this model, a given point $A$ is contained
within a bubble of size given by the largest surrounding spherical
region that contains enough ionizing sources to fully reionize itself.
If we ignore recombinations, then the ionized fraction in a region is
given by $\zeta f_{\rm coll}$, where $f_{\rm coll}$ is the collapse
fraction (i.e., the gas fraction in galactic halos) and $\zeta$ is the
overall efficiency factor, which is the number of ionizing photons
that escape from galactic halos per hydrogen atom (or ion) contained
in these halos. This simple version of the model remains valid even
with recombinations if the number of recombinations per hydrogen atom
in the IGM is treated as uniform; in this case, the resulting
reduction of the ionized fraction by a constant factor can be
incorporated into the value of $\zeta$.

In the extended Press-Schechter model [compare eq.~(\ref{eq:ePS})],
in a region containing a mass corresponding to variance $S_m$,
\beq f_{\rm coll} = {\rm erfc} \left(\frac{\del_c(z) - \del_m} {\sqrt{2 
(S_{\rm min} - S_m)}} \right) \ , \eeq where $S_{\rm min}$ is the
variance corresponding to the minimum mass of a halo that hosts a
galaxy, and $\del_m$ is the mean density fluctuation in the given
region. While this describes fluctuations in $f_{\rm coll}$ well, the
cosmic mean collapse fraction (and thus the overall evolution of
reionization with redshift) is better described by the halo mass
function of \citet{shetht99} (with the updated parameters suggested by
\citet{st02}). We thus use the latter mean mass function and adjust
$f_{\rm coll}$ in different regions in proportion to the extended
Press-Schechter formula; \citet{BLflucts} suggested this hybrid
prescription and showed that it fits a broad range of simulation
results. The resulting condition for having an ionized bubble of a
given size, written as a condition for $\del_m$ vs.\ $S_m$, is of the
same form as in \citet{fzh04}, at a given redshift, and thus (as they
showed) yields a linear barrier to a good approximation (see also
\citet{fmh06}).

In the model of \citet{fzh04}, the total fraction of points contained
within bubbles, as given by the model [i.e., eq.~(\ref{eq:Flin})],
comes out slightly different from the direct result for the mean
global ionized fraction, $x_i = \zeta f_{\rm coll}$ in terms of the
cosmic mean collapse fraction. To deal with this, we adopt the direct
values of $x_i$ versus redshift (or the values measured in a
simulation, when comparing to one), and adjust $\zeta$ within the
model to an effective value of $\zeta$ at each redshift that gives a
model value of $x_i$ that equals the desired one.

We illustrate the power of our solution from \S~4 by calculating a
number of different statistics. In the following examples we use the
cosmological parameters from \citet{Zahn} since we compare with their
results in the following subsection. In this subsection we assume that
the efficiency $\zeta$ is constant in all halos with circular velocity
$V_c$ above 16.5 km/s (corresponding to efficient atomic cooling);
letting reionization end, e.g., at $z=6.5$, yields a real $\zeta =
8.9$ (which is held fixed, independent of redshift, unlike the
effective model $\zeta$). Figure~1 shows the probability that two
points are both in ionized regions, divided (for visual clarity) by
the mean $x_i$, as a function of the distance between the points. The
probability is $F_>$ as given by eq.~(\ref{eq:F>}) evaluated at $S_1 =
S_2 = S_{\rm min}$. This probability in our model naturally satisfies
the limits $F_> \rightarrow x_i$ when $d \rightarrow 0$ (perfect
correlation) and $F_> \rightarrow x_i^2$ when $d \rightarrow \infty$
(no correlation), while these limits had to be artificially inserted
into previous models for the ionization correlation function. The
characteristic distance at which $F_>$ makes the transition between
these two limits grows as reionization proceeds, reflecting the
increase in the characteristic bubble size as larger and larger groups
of galaxies produce joint ionized regions.

\begin{figure}
\includegraphics[width=84mm]{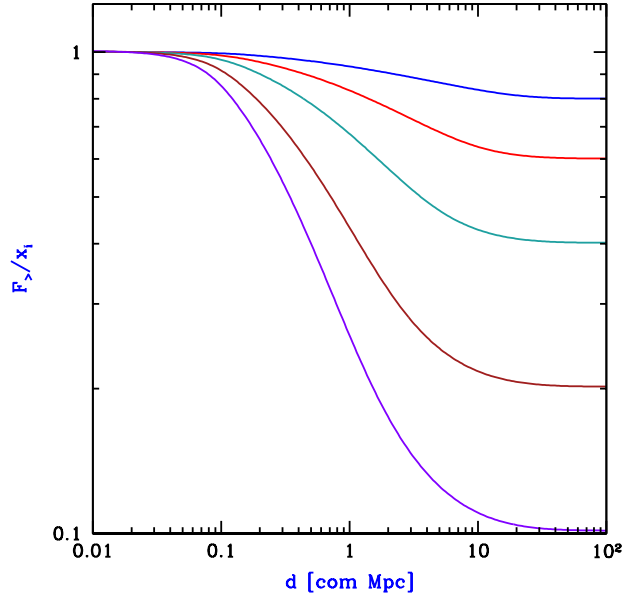}
\caption{Ionization correlations. We show the joint probability $F_>$ 
that two points separated by a distance $d$ are in ionized regions,
divided by the mean ionized fraction $x_i$. We consider $x_i=0.1$,
0.2, 0.4, 0.6 and 0.8 (bottom to top), assuming parameters for which
the universe fully reionizes at $z=6.5$ due to stars in halos with
efficient atomic cooling.}
\end{figure}

Our solution also allows us to calculate density-ionization
correlations. Figure~2 shows the probability distribution of the
density fluctuation on the scale $S_{\rm min}$ around a point $A$. In
this subsection and the next, we use the notation \beq \del_1(z_1)
\equiv D(z_1) \del_1\ , \label{eq:dOfz} \eeq where the growth factor
converts from the linearly-extrapolated $\del_1$ at redshift 0 (which
we have been using) to the linearly-extrapolated $\del_1$ at redshift
$z_1$. Given that point $A$ is neutral, we calculate as detailed in
\S~4.6 the separate probability distributions of $\del_1(z_1)$ given
that a point $B$ a distance $d$ away is either ionized or
neutral. When $d=1$ com Mpc the two points are highly correlated, and
point $B$ is most likely to be neutral as well, especially when
$\del_1(z_1)$ is very negative. However, when $d=10$ com Mpc the
correlations are weaker, and point $B$ is most likely ionized, but
only with a $65\%$ chance although the IGM as a whole is $80\%$
ionized in the plotted example.

\begin{figure}
\includegraphics[width=84mm]{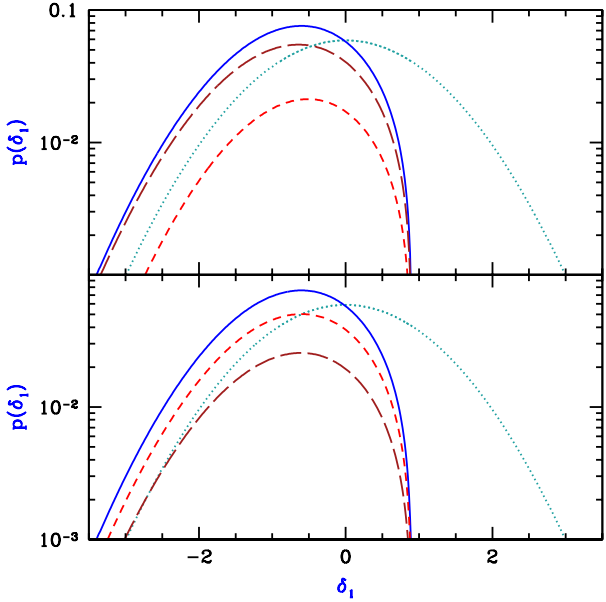}
\caption{Density-ionization correlations. We show in each panel the 
normalized probability distribution of the density fluctuation
$\delta_1(z_1)$ on the scale of 0.09 com Mpc (corresponding to $S_{\rm
min}=46$) around a point $A$, in general (dotted curve) or given that
point $A$ is in a neutral region (solid curve). For the latter, we
also show the break-down into a contribution from the case where point
$B$ is ionized (short-dashed curve) or where point $B$ is neutral
(long-dashed curve). With the same assumptions as in Figure~1, we
consider $x_i=0.8$ (which implies $z=7.3$) and a separation $d=1$ com
Mpc (top panel) or 10 com Mpc (bottom panel).}
\end{figure}

\subsection{The 21-cm Power Spectrum}

During cosmic reionization, we assume that there are sufficient
radiation backgrounds of X-rays and of Ly$\alpha$ photons so that the
cosmic gas has been heated to well above the cosmic microwave
background (CMB) temperature and the 21-cm level occupations have come
into equilibrium with the gas temperature. In this case, the observed
21-cm brightness temperature relative to the CMB is independent of the
spin temperature and, for our assumed cosmological parameters, is
given by \citep{Madau}
\beq T_b = \tilde{T}_b \Psi;\ \ \ \ \tilde{T}_b = 25 
\sqrt{\frac{1+z} {8}}\, {\rm mK}\ , \eeq with $\Psi = x_{\rm HI} 
[1+\delta(z)]$, where $x_{\rm HI}$ is the neutral hydrogen fraction
and we also used the notation of eq.~(\ref{eq:dOfz}). Under these
conditions, the 21-cm fluctuations are thus determined by fluctuations
in $\Psi$. To determine its statistical properties using our model, we
note that for a given random walk, the value of the neutral hydrogen
is either 1 (if the barrier has not been pierced) or 0 (if it has).

Consider points $A$ and $B$ at a distance $d$ from each
other. Then the correlation function of $\Psi$ is $\xi_{\Psi \Psi} =
\langle \Psi_1 \Psi_2 \rangle - \langle
\Psi_1 \rangle \langle \Psi_2 \rangle$, where the mean value is, e.g.,
for point 1:
\beq \langle \Psi_1 \rangle = F_{\rm <,lin}(\nu_1,\mu_1,S_1) + 
D(z_1)\, \bar{\delta}_{\rm lin}(\nu_1,\mu_1,S_1)\ ,
\eeq where we set $S_1 = S_{\rm min}$ and use eq.~(\ref{eq:dbar}).
This result arises from the fact that the average value of $\Psi$ is
simple the value of $[1+\delta(z)]$ averaged only within neutral
regions (where $x_{\rm HI}=1$, corresponding to random walks that have
not been absorbed); e.g., the first term (unity) yields simply the
fraction of the universe which is still neutral.

To average over the value of $[1+\delta_1(z_1)] \times
[1+\delta_2(z_2)]$ when both points are neutral, we follow the
derivation of eq.~(\ref{eq:F<}), writing, e.g., $(1+D(z_1)\, \del_1) =
(1+D(z_1)\, \del)+D(z_1)\, (\del_1-\del)$. Note that our solution for
$Q$ describes exactly those random walks that correspond to both
points being neutral (i.e., not absorbed by the barrier). We obtain
\beqa
\lefteqn{ \langle \Psi_1 \Psi_2 \rangle = \int_{\delta=-\infty}
^{\nu_{\rm m}+ \mu_{\rm m} \xi} d \del\ Q_{\rm lin}(\nu_{\rm
m},\mu_{\rm m},\del,\xi) } \nonumber \\ && \ \ \ \ \ \ \ 
\mbox{} \times \Bigl[ (1+D(z_1)\, \del) F_{\rm <,lin}(\nu_1+\mu_1 \xi - 
\del,\mu_1,S_1 - \xi) \nonumber \\ && \ \ \ \ \ \ \ \ \ \ \ \ \mbox{} 
+ D(z_1)\,\bar{\delta}_{\rm lin}(\nu_1+\mu_1 \xi - \del, \mu_1,S_1 -
\xi) \Bigl] \nonumber \\ && \ \ \ \ \ \ \ \mbox{} \times \Bigl[ (1+
D(z_2)\, \del) F_{\rm <,lin}(\nu_2+\mu_2 \xi - \del,\mu_2,S_2 - \xi)
\nonumber \\ && \ \ \ \ \ \ \ \ \ \ \ \ \mbox{} + D(z_2)\, \bar{\delta}
_{\rm lin}(\nu_2+\mu_2 \xi - \del, \mu_2,S_2 - \xi) \Bigl]\ .
\eeqa
We use this equation with $S_1 = S_2 = S_{\rm min}$. However, we can
calculate the correlation function of $\Psi$ down to smaller scales by
assuming that the smaller-scale power does not influence the
ionization and is independent of the larger scales. Thus, we simply
add to $\xi_{\Psi \Psi}$ the contribution $D(z_1) D(z_2) [
\xi_r(d,0,0) - \xi_r(d,r_{\rm min},r_{\rm min}) ]$ times the probability 
that both points are neutral, where $r_{\rm min}$ is the scale
corresponding to $S_{\rm min}$. Note, though, that these scales are
quite small and we expect large non-linear corrections to the model
over this range of scales.

Having calculated the correlation function of the 21-cm brightness
temperature, we Fourier transform it and obtain the power spectrum
$P(k)$ as a function of wavenumber $k$. We express the result in terms
of a characteristic quantity that has units of temperature: \beq
\Delta_{21}(k) \equiv \tilde{T}_b\, \sqrt{k^3 P(k)/(2 \pi^2)}\ . \eeq
Figure~3 shows that the 21-cm power spectrum changes shape during
reionization, acquiring large-scale power and flattening on scales up
to several tens of Mpc as the characteristic bubble size grows towards
the end of reionization. The amount of large-scale power depends
strongly on the bias of the typical ionizing galaxies. The bias is
larger for more massive halos, leading to stronger large-scale
fluctuations in this case. These trends are qualitatively similar to
those seen in previous approximate models that were constructed more
artificially [e.g., \citet{fmh06}].

\begin{figure}
\includegraphics[width=84mm]{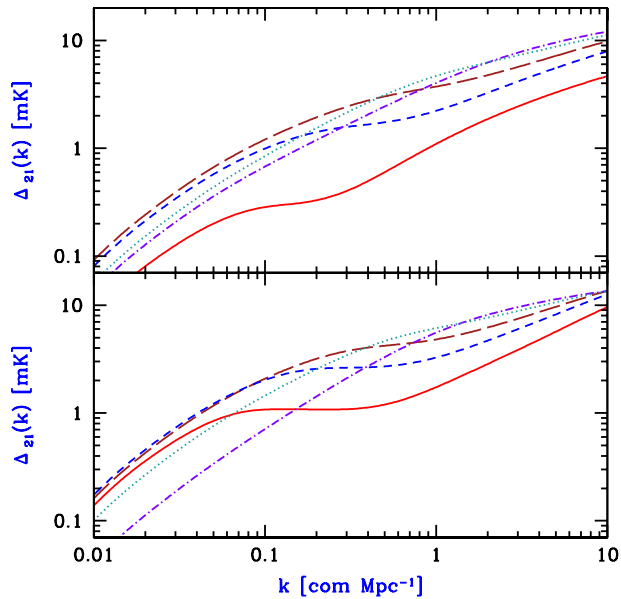}
\caption{21-cm power spectrum. Assuming parameters for which
the universe fully reionizes at $z=6.5$, we show in each panel results
for $x_i=0.1$ (dot-dashed curve), 0.3 (dotted curve), 0.5 (long-dashed
curve), 0.7 (short-dashed curve) and 0.9 (solid curve). We consider
stars forming in all halos above $V_c = 16.5$ km/s (top panel;
corresponds to efficient atomic cooling) or only in halos ten times as
massive, above $V_c = 35.5$ km/s (bottom panel; corresponds to strong
feedback in low-mass halos, e.g., due to photoheating or supernovae).}
\end{figure}

As a final example, in Figure~4 we compare our model quantitatively to
a numerical N-body plus radiative transfer simulation by
\citet{Zahn}. In this comparison we modify the model slightly in
accordance with the assumptions in their simulation. Unlike
\citet{fzh04}, who effectively assumed that the star formation rate in
halos is proportional to the rate of gas infall into them,
\citet{Zahn} assumed a constant mass-to-light ratio, which sets the
star formation rate in halos to be proportional to their total gas
content at a given time. This assumption leads to a slightly different
condition for having enough sources to ionize a given region [see
\citet{Zahn}], which we again approximate as a linear barrier 
constraint. We set the minimum halo mass to be $2 \times 10^9
M_{\odot}$, as assumed in the simulation, and set the effective
efficiency factor at each redshift so that the model yields the same
global ionized fraction as measured in the simulation. We also compare
our results to the numerical extended Press-Schechter model from
\citet{Zahn}, where they numerically applied the spherical ionization
condition (in real space) to the linear density field. 

These comparisons are an ambitious challenge for our model since it is
fully analytical and makes necessary approximations in using spherical
averages in the statistics, in applying simplifying assumptions that
are strictly valid only in $k$-space, and in neglecting significant
non-linear corrections. The model also relies on the two-step
approximation, approximates the reionization condition as a linear
barrier, and is based on a Lagrangian approach. The simulation is
limited as well, with fluctuations in the measured $\Delta_{21}(k)$
indicating a lack of convergence on large scales, while on small
scales the mass resolution corresponds to only 64 particles per $2
\times 10^9 M_{\odot}$ halo, well below the 500 required for
reasonable confidence as indicated by careful convergence tests
\citep{Springel03}. Nevertheless, the comparison indicates that the
simple analytical model captures the correct trends such as the change
in power-spectrum shape with redshift, and can therefore be used to
estimate the quantitative results and to explore the dependence on
model parameters such as the astrophysical properties of the ionizing
sources.

\begin{figure}
\includegraphics[width=84mm]{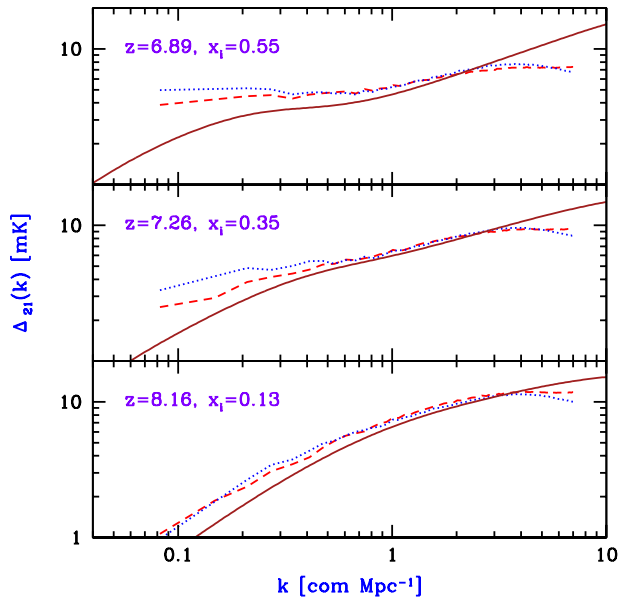}
\caption{21-cm power spectrum. We compare our model prediction (solid
curves) to those from the simulation (dashed curves) and numerical
extended Press-Schechter (dotted curves) from \citet{Zahn}. The
results are shown at several different redshifts, as indicated in each
panel. At each redshift $z$ we adjust the value of the efficiency in
our model in order to match the mean global ionized fraction $x_i$
from the simulation.}
\end{figure}

\section{Summary}

We have presented an approximate but fairly accurate analytical
solution to the mathematical problem of the joint evolution of two
correlated random walks with linear absorbing barriers. Our
self-consistent solution is a generalization of the constant-barrier
solution of \citet{sb} and is based on their two-step
approximation. Physically this mathematical setup can be applied to a
number of topics in galaxy formation where spatially-dependent
feedback or two-point correlations are important.

We have emphasized in particular the direct relevance to extended
Press-Schechter models of the ionizing bubble distribution during
cosmic reionization \citep{fzh04}. In this context, the joint
probability distribution $Q$ of the random-walk trajectories
[eq.~(\ref{eq:soln})] corresponds to the bivariate density
distribution at two points when both points are neutral. The bivariate
cumulative probability $F_>$ of both points hitting their barriers
[eq.~(\ref{eq:F>})] corresponds to the probability that both points
are in ionized regions. Other distributions [e.g., as given by
eq.~(\ref{eq:pdel1})] correspond to various elements of the joint
correlations among the densities and ionization states of the two
points.

We have shown that our model can be used not only to calculate
density-ionization correlations (Figures~1 and 2), but also (Figure~3)
the power spectrum of fluctuations in the 21-cm temperature
brightness, which may be observed in the next few years. Like any
analytical approach to complicated non-linear physics, our model is
approximate and simplified in a number of ways, but it correctly
captures the trends seen in simulations of reionization (Figure~4) and
thus can be used to explore various scenarios of cosmic reionization
and their observable consequences.

\section*{Acknowledgments}

The author is grateful for the kind hospitality of the {\it Institute
for Theory \& Computation (ITC)} at the Harvard-Smithsonian CfA, and
acknowledges support by Harvard university, Israel Science Foundation
grant 629/05 and Israel - U.S. Binational Science Foundation grant
2004386.

\bsp

\label{lastpage}

\end{document}